# Calibration and electric characterization of p-MNOS RADFETs at different dose rates and temperatures

P. A. Zimin[a,b], E. V. Mrozovskaya[a,b], P. A. Chubunov[a,b], V. S. Anashin[a], G. I. Zebrev[a,b*]

[a] JSC Institute of Space Device Engineering, Moscow, Russia
[b] National Research Nuclear University MEPHI, Moscow, Russia

**Abstract:** This paper describes the radiation response and I-V characteristics of the stacked p-MNOS based RADFETs measured at different dose rates and irradiation temperatures. It is shown that the enhanced charge trapping takes place at the interface of the thick gate dielectrics in the MNOS transistors at low dose rates (ELDRS). The sensitivity of the radiation effect to irradiation temperature has also experimentally revealed. We associate both effects with the temperature and dose rate dependence of the effective charge yield in the thick oxides described within the framework of the previously proposed model. We have also simulated the I-V characteristics of the transistors for different total doses and irradiation conditions. It has been found the used electric and radiation models qualitatively and semi-quantitatively describe the observed dependencies of the RADFETs' sensitivity on dose rates and irradiation temperatures for the devices with different thickness of insulators.

*Keywords:* RADFET; ELDRS; Total dose effects; Dose rate effects; Simulation; Dosimeter;

## 1. Introduction

The field-effect-transistor-based dosimeters in which the threshold voltage shift is a quantitative indicator of the absorbed dose are now widely used due to their simplicity, low cost, low power consumption, and compatibility with the CMOS technology [1]. The devices used in dosimetry are typically the p-channel MOS or p-MNOS (Metal – Nitride – Oxide - Semiconductor) [2, 3] transistors (RADFETs). The design of the RADFETs encounters with the challenges in ensuring of stability and reproducibility of their radiation and electrical characteristics at different radiation and environmental conditions. Particularly, the I-V characteristics of the MOSFETs are temperature-dependent which makes it difficult to use the RADFETs in a wide range of ambient temperatures. Besides, one of the significant challenges is that the RADFET based dosimeters may be dose rate sensitive. To enhance the sensitivity of the RADFETs, they are usually made with rather thick gate dielectrics (typically > 100 nm). This could lead to the occurrence of the enhanced low dose rate sensitivity (ELDRS), which hinders the calibration of dosimeters at moderate high dose rates. To the best of our knowledge, this effect in the MOSFET based dosimeters was observed for the first time in [4], and it has not been yet properly investigated experimentally and theoretically (see, also [5, 6]). The ELDRS is also temperature-dependent effect which alters the temperature response of the MOSFET's electric characteristics. Generally speaking, the development of the RADFET dosimeters requires a combined investiga-tion of their electrical and radiation characteristics under different conditions. This work aims to investigate and simulate the radiation response of the stacked MNOS based p-RADFETs at different dose rates and irradiation temperatures.

## 2. Experiments and modeling

### 2.1. Experimental conditions

We investigated two types of the stacked MNOS-based devices (at least 4 samples for each type). The thickness of the $Si_3N_4$ layer in both MNOS devices was 150 nm, whereas the thicknesses of the $SiO_2$ layers were 150 nm in the "thin" devices and 500 nm in the "thick" devices (see Fig. 1). The channel width-to-length ratio W/L for the "thick" device was approximately three times more than for the "thin" device.

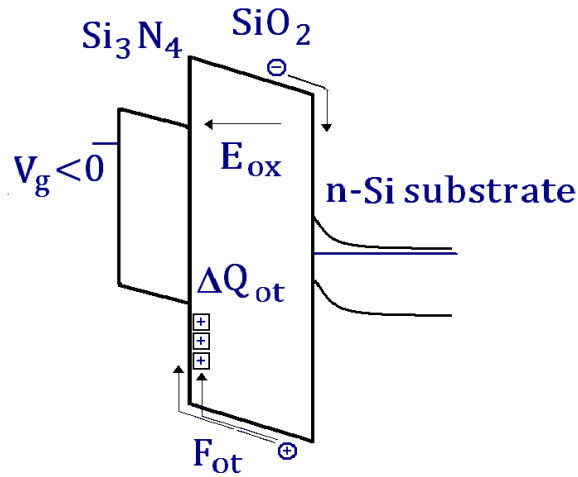

Fig. 1. The band diagram of the p-MNOS based RADFET.

The MNOS devices were irradiated with different dose rates from 0.01 Gy/s to 1 Gy/s using a Co-60 source. Irradiation and measurements were performed at different temperatures (-40°C, +25°C, and +60°C). The experimental dose error did not exceed 20%. The measurement circuit is shown in Fig. 2. The threshold voltage shift during irradiation was tracked at a fixed output drain current. The reference drain current $I_D^{(r)}$ and the gate biases were approximately chosen at the "Zero Temperature Coefficient (ZTC) point" [7]. All measurements were performed with a drain bias $V_D = -0.2$ V. The backward substrate bias $V_{sub} = +2.0$ V was applied to expand the dynamic range of the measurements. The operation gate voltage range was from 0 to - 12 Volts.

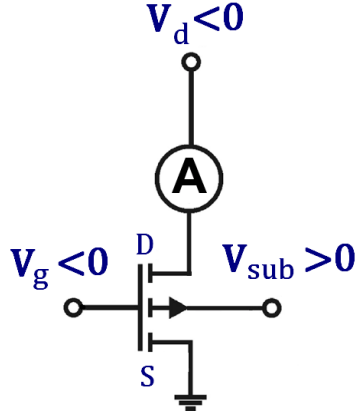

Fig. 2. The measurement electric scheme, $V_D$ = - 0.2 V, $V_{sub}$ = 2.0 V. The gate bias during the irradiation $V_G$ was changed for a fixed reference current in the range from - 1.5V to – 12 V.

The "thick" and the "thin" MNOS transistors have different ZTC points, and therefore different reference drain currents $I_D^{(r)}$ (145 μA for the "thick" and 55 μA for the "thin" samples).

## 2.2. Modeling of p-MNOS based RADFETs

The radiation-induced charge per unit area $\Delta Q_{ot}$, trapped at the Si$_3$N$_4$-SiO$_2$ interface, can be estimated as follows

$$\Delta Q_{ot} = q\, \eta_{eff} \left( E_{ox}, P, T \right) F_{ot} K_g t_{ox} D \,,\tag{1}$$

where $D$ is total dose, $q$ is the electron charge, $F_{ot}$ is the dimensionless hole trapping efficiency at the Si$_3$N$_4$–SiO$_2$ interface, $K_g \cong 8{\times}10^{14}$ cm$^{-3}$/Gy is the electron-hole pair generation rate constant in SiO$_2$, $t_{ox}$ is the silicon oxide thickness, and $\eta_{eff}\left( E_{ox}, P, T \right)$ is the effective charge yield dependent on the oxide electric field $E_{ox}$, dose rate $P$, and irradiation temperature $T$ (see Appendix). The charge yield is a very important parameter determining the charge trapping in the MOS devices at different conditions [8]. The threshold (reference) voltage shift due to the charge trapping at the Si$_3$N$_4$-SiO$_2$ interface can be described as follows

$$\Delta V_T = -\frac{\Delta Q_{ot} t_N}{\varepsilon_0 \varepsilon_N} \,,\tag{2}$$

where $t_N$ is the Si$_3$N$_4$ layer thickness, and $\varepsilon_0 \varepsilon_N$ is the silicon nitride dielectric permittivity ($\varepsilon_N \cong 7.5$). Then, the sensitivity of an MNOS RADFET can be naturally defined as

$$S = \frac{\left| \Delta V_T \right|}{\Delta D} = q\frac{\eta_{eff}\left( E_{ox}, P, T \right) F_{ot} K_g t_{ox} t_N}{\varepsilon_0 \varepsilon_N} \,.\tag{3}$$

The parameters $F_{ot}$ and $\eta_{eff}$ were used to fit the sensitivity $S$ for relatively low-dose-rate irradiation (~$10^{-2}$ Gy/s) at room temperature followed by a fitting with $\eta_{eff}$ for other dose rates and temperatures. We will show below that the effective charge yield $\eta_{eff}$, simulated with the model, described in the Appendix, allows a consistent description for different dose rates (ELDRS) and irradiation temperatures.

### *2.3. Calibration tests of the dosimeters*

The calibration tests were performed at the different dose rates and irradiation temperatures. Some results of the calibration measurements are shown in Fig. 3. All the curves exhibit approximately linear (more precisely, slightly sublinear) dose behavior without noticeable saturation in the working range of doses determined by the operation range of the output threshold voltage. Presented results show a pronounced enhancement of the charge trapping sensitivity at the low dose rates (ELDRS) [9]. We attributed such behavior to the dependence of the charge yield on dose rate. The physical model of this effect has been developed and validated in a series of the works [10, 11, 12, 13]. A synopsis of this model is presented in the Appendix.

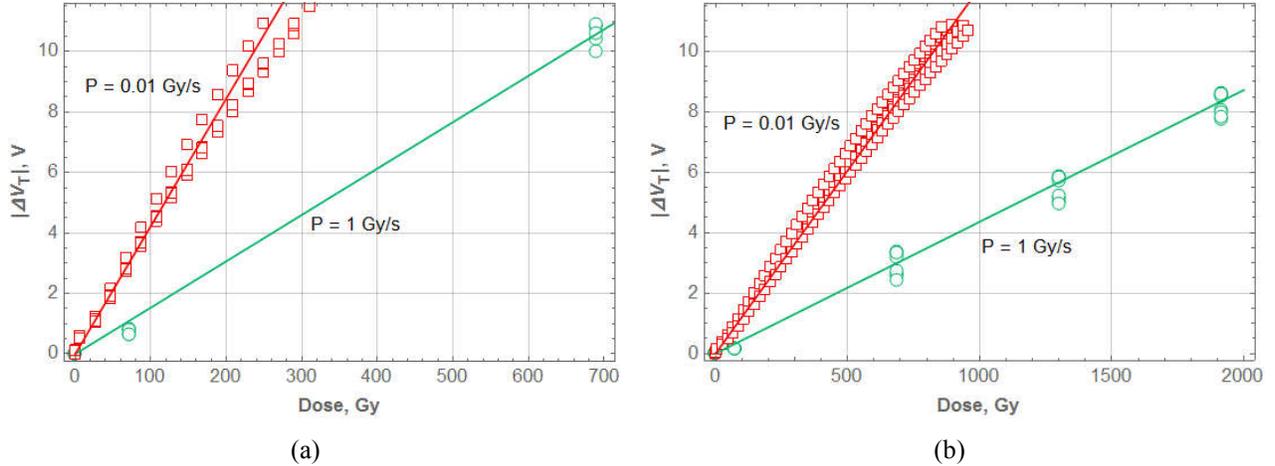

Fig. 3. Comparison of the experimental (points) and simulated (lines) dose dependencies of the threshold voltage shift at different dose rates $P$ = 1 Gy/s (circles) and $P$ = 0.01 Gy/s (squares) for the "thick" (a) and "thin" (b) devices, irradiated at room temperature T = +25 °C.

Using Eqs. 1 - 3 and A1 - A3, we fitted the parameters of the radiation model, taking into account the real geometric sizes of the samples. Comparison of measurements and simulations in Fig. 3 shows a good agreement between the charge yield model predictions and experimental results at different dose rates. The radiation parameters of the model are listed in Table A1 of the Appendix.

### 2.4. Electric characterization before irradiation

The I-V characteristics for all samples prior irradiation were measured at three different temperatures (-40°C, 0°C, +50°C). The electric characterization of the transistors was carried out on the basis of the MOSFET model [14]. Figure 4 shows a comparison between experimental and simulated results.

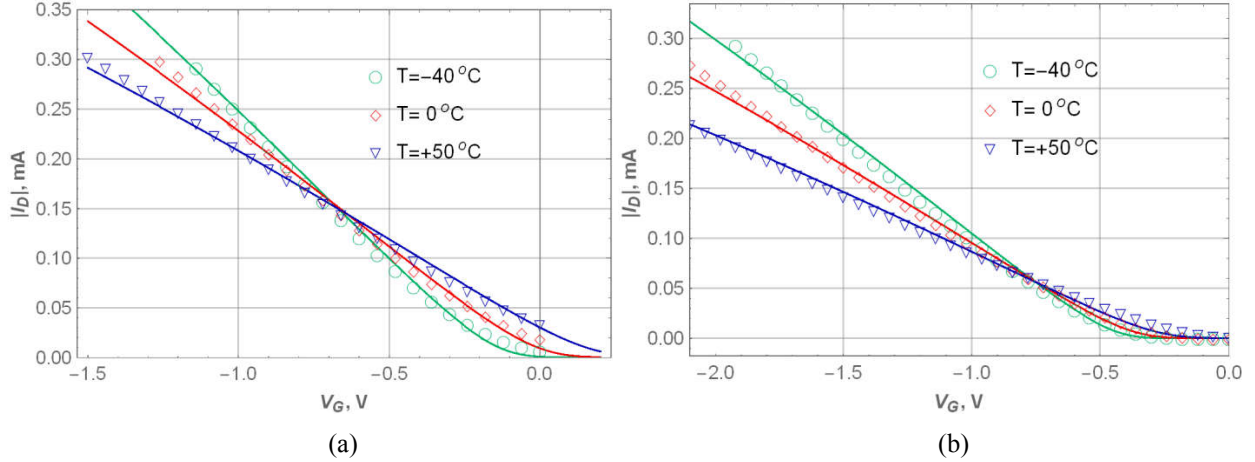

<div align="center">(a)　　　　　　　　　　　　(b)</div>

Fig. 4  Experimental (tokens) and simulated (lines) I–V results at different temperatures for the "thick" (a) and "thin" devices (b). The parameters fitted in the characterization were used further for the simulation of dose dependencies.

The input and fitted parameters of the unirradiated transistors are listed in Table I.

*Table I. Electrical model parameters*

| Devices\parameters | $t_N$, nm | $t_{ox}$, nm | $W/L$ | Channel hole mobility $\mu_0$, cm²/V s (fitted) | $\alpha$, mV/K (fitted) | $V_{T0}$, V (fitted) |
|---|---|---|---|---|---|---|
| "Thick" devices | 150 | 500 | 580 | 320 | 3.5 | 0.2 |
| "Thin" devices | 150 | 150 | 150 | 350 | 2.0 | -0.2 |

The geometric sizes ($t_N$, $t_{ox}$, $W/L$) were known from the technical characteristics of the test structures. The temperature dependence of the threshold voltage of the p-MOSFETs in the range 200-400K can be described with the temperature coefficient $\alpha$ as follows

$$V_T(T) = V_{T0} + \alpha(T - T_0),\tag{4}$$

where $T_0$ is the reference temperature which was set to be equal to 300 K. The temperature dependence of the channel hole mobility $\mu_0(T)$ in strong inversion is mainly determined by the phonon scattering and can be simulated as follows

$$\mu_0(T) = \mu_0\left(\frac{T_0}{T}\right)^{3/2},\tag{5}$$

where $\mu_0$ is the mobility at the reference temperature. The fitted parameters of the electrical model ($\alpha$, $V_{T0}$ and $\mu_0$) were assumed to be dose-independent and were then used for simulation of the irradiated devices.

## 3. Simulation of I-V characteristics at different temperatures and doses

The I-V curves measured for different total doses in the devices irradiated with a dose rate of $1.9 \times 10^{-2}$ Gy/s at the elevated and low temperatures are shown in Fig. 5. The shifts of the I-V curves were simulated for different doses with the model parameters, presented in Table A1.

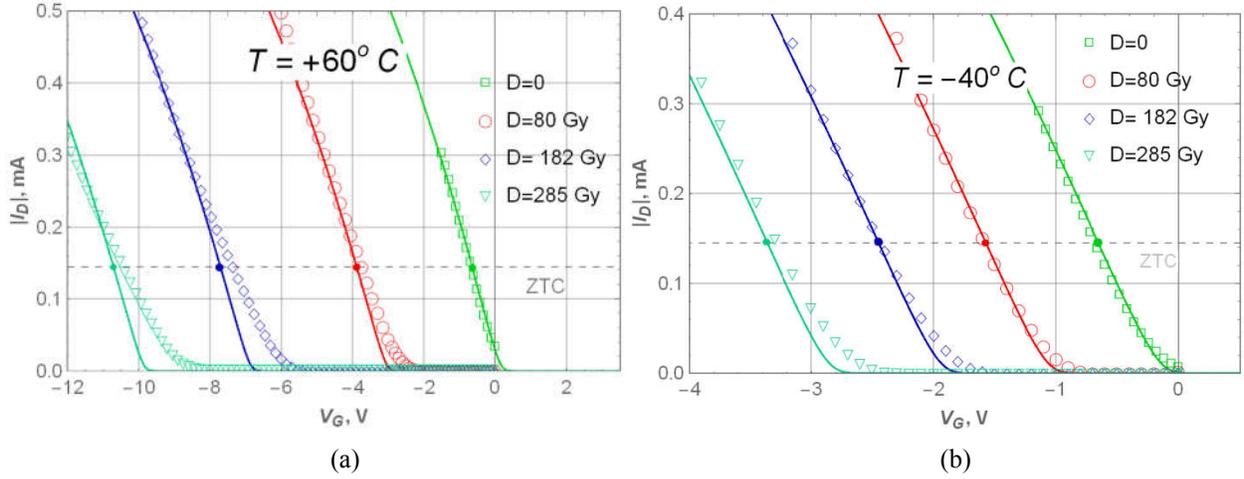

(a)                                                                (b)

Fig. 5. Comparison of experimental (points) and simulated (lines) I-V curves in the "thick" devices ($t_{ox}$ = 500 nm) for the elevated (a) $T$ = +60°C and for the low (b) $T$ = -40°C temperatures, P = $1.9 \times 10^{-2}$ Gy /s. The ZTC current is marked by the dashed lines.

The transconductance at the large gate overdrive biases was approximately dose-independent. The change in the shape of the I-V curves in the weak inversion and in the subthreshold regimes is due to the radiation-induced inter-face trap buildup.

The values of the effective charge yield $\eta_{eff}$ were found for each experimental I-V curve. Figure 6a shows the dose dependencies of the effective charge yield values $\eta_{eff}$ fitted from the results in Fig.5.

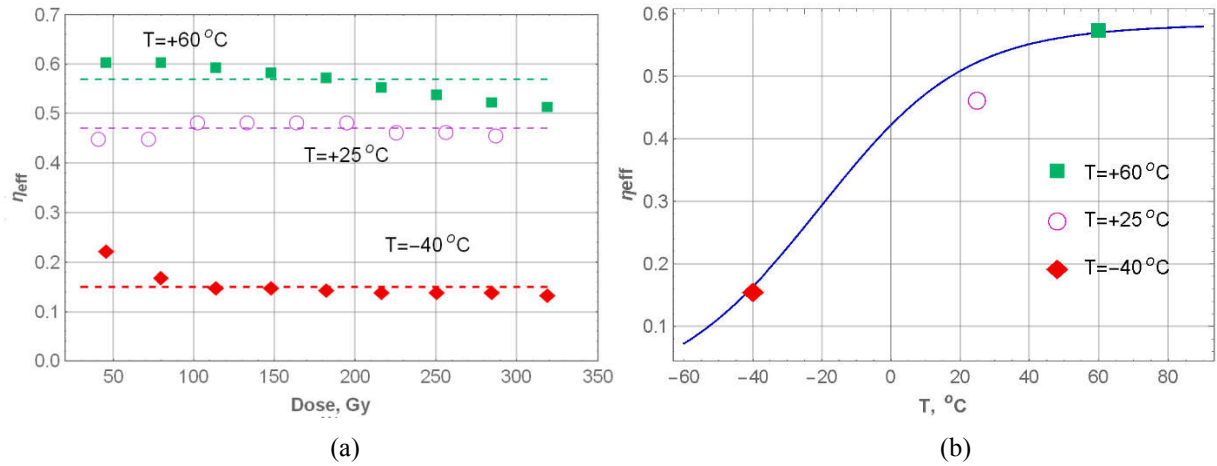

(a)                                                                (b)

Fig. 6. (a) The effective charge yield $\eta_{eff}$ fitted from results in Fig. 5 as functions of a dose at three different temperatures, the dashed lines marked the simulation results;  (b) the temperature dependence of $\eta_{eff}$ simulated with (A1-A3) for $P$ = $1.9 \times 10^{-2}$ Gy/s.

The charge yield $\eta_{eff}$ practically does not depend on dose, but strongly depends on the irradiation temperature in full consistency with the charge yield model, described in the Appendix. The comparison between the simulation and the experimental points at different irradiation temperatures is shown in Fig. 6b.

Fig. 7a shows the dose dependencies of the charge yield fitted for different dose rates, which explicitly exhibit the ELDRS effect. Fig. 7b presents a comparison of the theoretical dependence of $\eta_{eff}$ on the dose rate (see equations (A1)-(A3) ) and the values obtained from the experimental data.

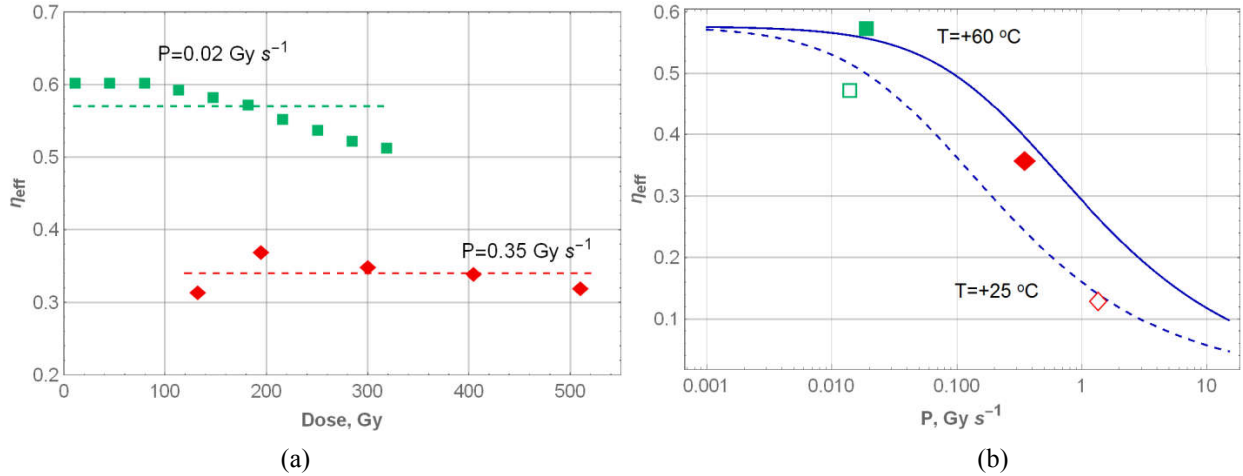

(a)                                    (b)

Fig. 7. (a) The effective charge yield $\eta_{eff}$ in the "thick" (500 nm) devices as functions of dose with different dose rates; (b) comparison between experiment and simulation for $T = +60°C$ (filled tokens and solid line) and $T = +25°C$ (blank tokens and dashed line).

Similar electric and radiation characterization was performed for the "thin" ($t_{ox} = 150$ nm) devices. Fig. 8 shows the experimental and simulated charge yield $\eta_{eff}$ in the "thin" devices irradiated with a dose rate $\sim 2 \times 10^{-2}$ Gy/s at different temperatures.

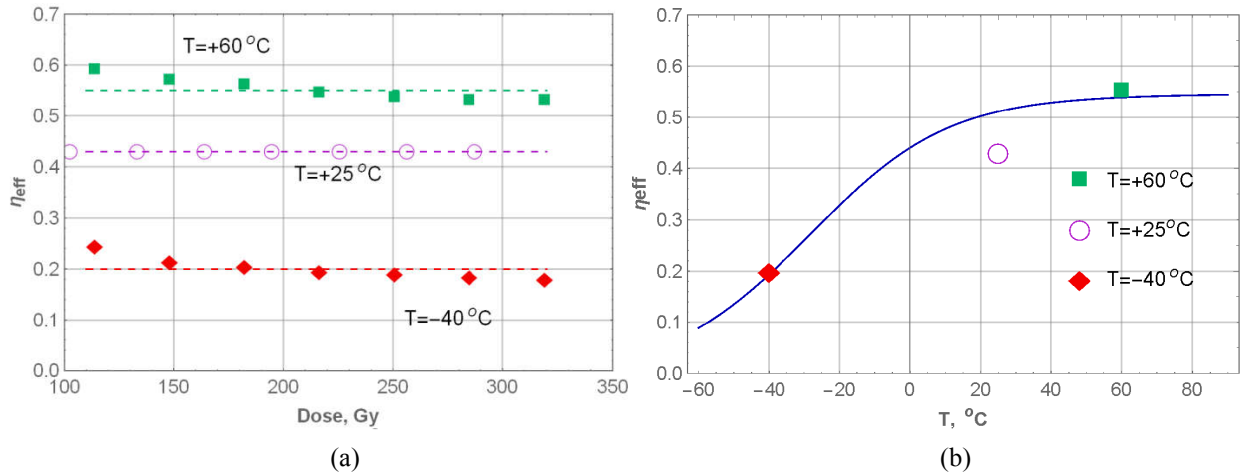

(a)                                    (b)

Fig. 8. The effective charge yield $\eta_{eff}$ in the "thin" (150 nm) devices as functions of dose under irradiation with a dose rate $1.9 \times 10^{-2}$ Gy/s at different temperatures; (b) comparison between the experiments and simulation.

Finally, Fig. 9 shows the experimental and the simulated charge yield in the "thin" devices irradiated at $T = +60°C$ with different dose rates.

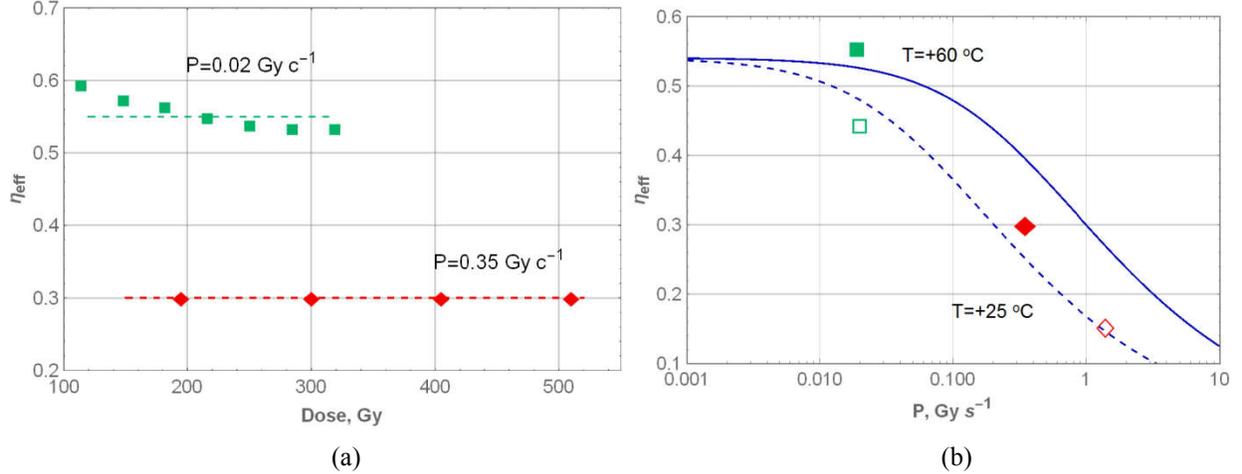

<center>(a)</center>

<center>(b)</center>

Fig. 9. The effective charge yield $\eta_{eff}$ in the "thin" (150 nm) devices as functions of dose for irradiation at $T = +60°C$ with different dose rates; (b) comparison between the experiments and simulation for $T = +60°C$ (filled tokens and solid lines) and $T = +25°C$ (blank tokens and dashed lines).

The devices of both types exhibit at elevated temperatures $(T = +60°C)$ and relatively low dose rates $(P = 1.9×10^{-2}°Gy/s)$ a noticeable decrease of the formally fitted $\eta_{eff}$ as the dose increases. We associate this effect with an annealing process which has not yet been taken into account in our simulations. We found, that the annealing ("fading" of the dosimeters) during irradiation in our devices has a form of an approximately logarithmic temporal dependence [15]. This issue requires further experimental and theoretical investigations.

## 4. Summary

This paper provides a detailed investigation of electrical characteristics of the stacked p-MNOS transistors irradiated with the different dose rates at different temperatures. We have provided a convincing experimental evidence of the ELDRS effects in the thick oxides of the p-MNOS based RADFETs. This is critically important in such devices since the ELDRS effects cause an extremely unwanted sensitivity of the dosimeters to the variable dose rates in space. It was shown that the proposed model can describe both qualitatively and semi-quantitatively the observed dependencies of the dosimeter sensitivity $S$ on dose rate and irradiation temperature. This allows correcting the p-MNOS based dosimeter indications based on a preliminary characterization of the p-MNOS transistor. We believe that the results presented may be more or less common to a wide class of the field-effect-transistor-based sensors including new emergent devices [16, 17].

## Appendix: Modeling ELDRS in the thick oxides

The enhanced low dose rate sensitivity is often associated only with bipolar devices. We have been assuming in [10, 11] that the ELDRS is a general property of the thick amorphous insulators with low internal electric field. We also supposed that the ELDRS effects may also occur in the thick isolation oxides of the MOS devices. In particular, the enhanced trap-assisted electron-hole recombination at relatively high dose rates may result in reducing of the effec-tive charge yield in the thick oxides. It was found, that the effective charge yield $\eta_{\text{eff}}$, limited by recombination between the mobile electrons and the holes, localized at shallow bulk traps, can be described as a follows

$$\eta_{\text{eff}}\left(P,T,t_{\text{ox}},E_{\text{ox}}\right)=\eta\left(E_{\text{ox}}\right)\frac{\left(1+4f\right)^{1/2}-1}{2f},\tag{A1}$$

$$f\left(E_{\text{ox}},P,T\right)\cong\frac{q\,t_{\text{ox}}^{2}}{6\varepsilon_{\text{ox}}\varepsilon_{0}\,\mu_{\text{p}}\,E_{\text{ox}}^{2}}\,\eta\left(E_{\text{ox}}\right)K_{\text{g}}\,P\exp\!\left(\frac{\varepsilon_{\text{p}}}{k_{\text{B}}T}\right),\tag{A2}$$

where $\mu_{\text{p}}$ is the hole mobility in the $SiO_2$, $T$ is the irradiation temperature, $k_{\text{B}}$ is the Boltzmann constant. The high-field part of the charge yield dependence $\eta\left(E_{\text{ox}}\right)$ is normally interpolated by a monotonically increasing function of the internal electric field $E_{\text{ox}}$ [18, 19]

$$\eta\left(E_{\text{ox}}\right)=\eta_{0}+\frac{E_{\text{ox}}/E_{0}}{1+E_{\text{ox}}/E_{0}}\left(1-\eta_{0}\right),\tag{A3}$$

where $\eta_{0}$ and $E_{0}$ are the fitting constants, parameterizing a high-field region (when $f\ll1$) of the experimental dependence. The temperature dependence of $\eta_{\text{eff}}$ is determined by the effective energy depth of the shallow hole traps in the bulk oxide $\varepsilon_{\text{p}}$ (see Fig. 1A [20]).

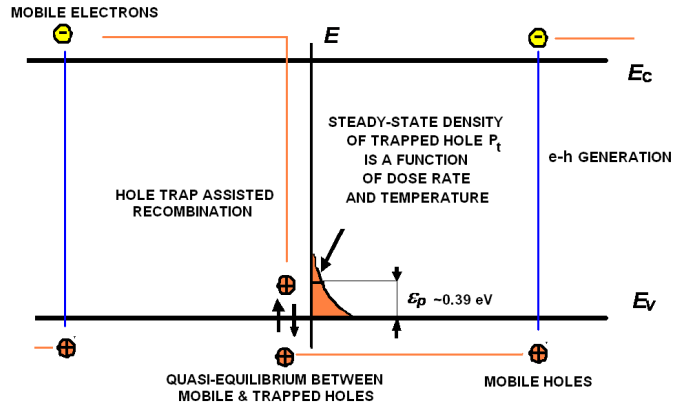

Fig. A1. Mechanism of electron-hole recombination through the hole levels localized near the valence band edge of the $SiO_2$.

All parameters of the model used for simulations are presented in Table A1. In fact, only $E_{\text{ox}}$ and $F_{\text{ot}}$ were varied as fitting parameters in Table A1. The basic physical parameter that determines the temperature dependence of the charge yield experimentally found in [11] ($\varepsilon_{\text{p}}\cong0.39$ eV) has been successfully used to describe radiation-induced degradation in different devices [10-13, 20].

*Table A1. Radiation model parameters*

|  | $F_{ot}$ | $E_{ox}$, V/cm | $\varepsilon_p$, eV | $\eta_0$ | $E_0$, V/cm | the oxide hole mobility cm$^2$/(V s), $\mu_p$ |
|---|---|---|---|---|---|---|
| "Thick" devices | 0.47 | $8\times10^4$ | 0.39 | 0.52 | $4.5\times10^5$ | $10^{-5}$ |
| "Thin" devices | 0.49 | $2.3\times10^4$ | 0.39 | 0.52 | $4.5\times10^5$ | $10^{-5}$ |

**Acknowledgment**

This work was supported by the Competitiveness Program of the NRNU MEPHI, Russia.